\documentstyle[prl,aps,multicol]{revtex}

\begin{document}

\title{The analysis of Hardy's experiment revisited}

\author{ Lev Vaidman} 

\address{School of Physics and Astronomy, 
Raymond and Beverly Sackler Faculty of Exact Sciences, \\
Tel-Aviv University, Tel-Aviv 69978, Israel.} 

\date{}

\maketitle

\begin{abstract}
  Cohen and Hiley [ Phys. Rev. {\bf A 52}, 76 (1995)] have criticized
  the analysis of Hardy's gedanken experiment according to which
  the contradiction with quantum theory in Hardy's experiment arises due
  to the failure of
  the ``product rule'' for the elements of reality of pre- and
  post-selected systems. It is argued that the criticism of Cohen and
  Hiley is not sound.
\end{abstract}

\begin{multicols}{2}
 
  Cohen and Hiley\cite{CoHi} analyzed the
discussions\cite{BG,CN,Ha94,V-prl} which followed a
  gedanken experiment proposed by Hardy\cite{Ha}. In his original
  letter Hardy claimed that this gedanken experiment proves the
  impossibility of construction of Lorentz-invariant elements of
  reality. I pointed out\cite{V-prl} that Hardy's proof relies on the
  ``product rule'' of elements of reality. This rule  does not hold for pre-
  and post-selected quantum systems  considered in
  Hardy's experiment. Cohen and Hiley, however, claimed that the
  contradiction in the analysis of Hardy's experiment is a consequence
  of the well-known noncovariance of the reduction postulate. They
  claimed that the contradiction  has
  nothing to do with realism and that the non-applicability of the
  product rule is essentially unrelated to this (the detailed analysis
  of the latter issue  appears in Ref.\cite{CoHi-Foun}).

   Based on Cohen\cite{Co},  Cohen and Hiley also claimed  that my
  analysis of Hardy's example is not valid because it ``makes
  incorrect use of the formula of Aharonov, Bergmann and Lebowitz''
  \cite{ABL}. I show that Cohen's argument is not sound
  elsewhere\cite{V-com-Co,V-coun}.

  In this comment I will argue that, contrary to the Cohen and Hiley
  claims,  the arguments of Hardy went beyond showing
  the noncovariance of the projection postulate and that the
  contradictions  obtained on the basis of Hardy's
  experiment do not hold if we do not accept the product rule or the
  closely connected ``and rule''\cite{V-er}.

A  covariant reduction of  quantum states (i.e.  collapse)
is not a prerequisite for a  Lorentz-invariant theory  consistent with
quantum mechanics. Indeed, there are several proposed interpretations
of quantum theory  without
collapse: the Many-Worlds Interpretation\cite{E-mwi}, various
 hidden variable theories such as the Causal
Interpretation\cite{Bohm}, and
some constructions based on the two-state vector
formalism\cite{Va-Ph,AR}.  Hardy did not consider the many-worlds
option and he overlooked the possibilities of
the two-state vector approach, but he did consider
  hidden variable theories. Clearly, the noncovariance of the collapse
does not affect the Lorentz-invariance of such (non-collapse)  theories.
Therefore, even if the noncovariance of the collapse is established, the
question of whether a Lorentz-invariant quantum theories exist remains
open.  Hardy's work was a step towards answering this question.

In order to make a comparison between    the arguments demonstrating
the non-covariance of the
 collapse  and Hardy's argument I will analyze a
 simple example. 
 Following Aharonov and Albert\cite{AA}, consider a particle located in
three separate boxes $A, B$ and $C$. The particle is prepared in the
initial state
\begin{equation}
 {1\over\sqrt 3} (|A\rangle +|B\rangle +|C\rangle),
\end{equation}
 where  $|A\rangle$ signifies a particle located in box $A$, etc. Assume
that in one Lorentz frame boxes $A$ and $C$ are opened simultaneously
and the particle is found in $C$. Then, in a Lorentz frame in which
the measurement in box $A$ is performed first, the evolution of the
state is:
\begin{equation}
{1\over\sqrt 3} (|A\rangle +|B\rangle
+|C\rangle) ~ \rightarrow ~ {1\over\sqrt 2} (|B\rangle
+|C\rangle) ~ \rightarrow  ~|C\rangle .
\end{equation}
However, in a Lorentz frame in which box $C$ is opened first, the
evolution is: 
\begin{equation}
{1\over\sqrt 3} (|A\rangle +|B\rangle
+|C\rangle)~ \rightarrow ~|C\rangle .
\end{equation}
Therefore, in one frame the particle was located in two boxes for some
period of time while in another it was never located in two boxes
(it was located in all three and then just in one). This radical
difference in the descriptions of the two Lorentz observers indicates
the noncovariance of the collapse.

Let us turn now to Hardy's argument.
In order to make the comparison more clear  
I will
consider the analog of    Hardy's original proposal 
for 
 a system of
two spatially separated  spin-1/2 particles. Assume that  the two
particles are prepared 
in the initial state (in the spin $z$ representation)
\begin{equation}
|\Psi_1\rangle ={1\over \sqrt 3}(~|{\uparrow}\rangle_1 |{\uparrow}\rangle_2 +
|{\downarrow}\rangle_1 |{\uparrow}\rangle_2
+|{\uparrow}\rangle_1 |{\downarrow}\rangle_2 ~) .
\end{equation}
Then, in a given Lorentz frame, simultaneous measurements of both spins
in $x$ direction
 are performed and found to be ``down'', i.e. the system is found in the
state \begin{equation} |\Psi_2\rangle ={1\over 2} (|{\uparrow}\rangle_1 -
|{\downarrow}\rangle_1)(|{\uparrow}\rangle_2-|{\downarrow}\rangle_2). 
\end{equation} (Since $\langle \Psi_2|\Psi_1\rangle \neq 0$ this result is
possible.)  Straightforward calculations show that in this situation the
Lorentz observer who sees the measurement of particle 1 being performed
first concludes that after this measurement and before the measurement
performed on particle 2 the state of particle 2 is
$|{\downarrow}\rangle_2$. Similarly, the observer who sees the measurement
on particle 2 being performed first concludes that the state of particle 1
after this and before the measurement performed on particle 1 is
$|{\downarrow}\rangle_1$.  The time evolutions of the state of the two
particles according to the two observers are:  \begin{eqnarray}
|\Psi_1\rangle \rightarrow {1\over {\sqrt 2}} (|{\uparrow}\rangle_1 -
|{\downarrow}\rangle_1)|{\downarrow}\rangle_2 \rightarrow |\Psi_2\rangle
,\\ |\Psi_1\rangle \rightarrow {1\over {\sqrt 2}}
|{\downarrow}\rangle_1(|{\uparrow}\rangle_2-|{\downarrow}\rangle_2)\rightarrow
|\Psi_2\rangle .  \end{eqnarray} The descriptions of the two observers are
different, but not so radically different as in the previous example.
Anyway, this difference was not used as a basis of the arguments leading
to contradiction in Hardy's experiment. 

 In order to reach the contradiction, Cohen and Hiley {\em combined} the
statements of one Lorentz observer about particle 1 and the other observer
about the state of particle 2 and concluded that the state of the
two-particle system before the measurements of the $x$ spin components was
\begin{equation} |\Psi \rangle = |{\downarrow}\rangle_1
|{\downarrow}\rangle_2.  \end{equation}
 However, the initial state is orthogonal to
this state, $\langle \Psi |\Psi_1\rangle = 0$,  and this is the
contradiction of Cohen and Hiley.\cite{foot1}
 
In the process of combining the statements of the two Lorentz
observers Cohen and Hiley used a variation of the ``and
rule''\cite{V-er}: if $A=a$ is an element of reality and $B=b$ is an
element of reality then \{$A=a$ and $B=b$\} is also an element of
reality. In this case ``$A=a$'' is replaced by the statement about the
state of particle 1,   $|\Psi\rangle_1 = |{\downarrow} \rangle_1$, and
 ``$B=b$'' is replaced by  $|\Psi\rangle_2 = |{\downarrow} \rangle_2$. The
 ``and rule'' is closely connected to the product rule and it also
 {\em does not hold} for the pre- and post-selected
 systems\cite{V-er}. Thus, it is not
surprising that adopting the ``and rule''  leads to a contradiction in the
analysis of  Hardy's experiment.

Beyond showing that the core of these contradictions lies in the failure
of the product rule and the ``and rule'', it is important to analyze the
possibility of relaxing the requirement that these rules are fulfilled
(thus allowing the construction of Lorentz-invariant elements of reality
for pre- and post-selected quantum systems\cite{V-prl}). I argue that the
failure of the ``and rule'' is indeed what happens in Hardy's experiment. 
The physical (operational) meaning of the statement that the state of
particle 1 is $|\Psi\rangle_1 = |{\downarrow} \rangle_1$ is: a measurement
of the $x$ spin component must yield $\sigma_{1x} = -1$. This statement
must be supplemented by the following condition: ``provided we do nothing
which might disturb the state of the particle''. In Hardy's experiment
this condition means, in particular, that we do not make a measurement of
the $x$ spin component of particle 2. Similarly, a measurement of the $x$
spin component of particle 2 must yield $\sigma_{2x} = -1$, provided we do
not measure the spin
 of particle 1.  Obviously, combining this statements does
not tell us what will be the results of the   $x$ spin component measurements
performed on both particles together.

Abandoning the ``and rule'' allows us to provide a Lorentz-invariant
description of a pre- and post-selected quantum system\cite{Va-Ph,AR}.
Although we cannot combine the statements $|\Psi\rangle_1 =
|{\downarrow} \rangle_1$ and $|\Psi\rangle_2 = |{\downarrow}
\rangle_2$ in a naive way, we can find a certain operational meaning
of these two statements together. It has been shown\cite{AV91} that in
any situation in which the outcome of a standard measurement of a
 variable is known with certainty, the outcome of the {\em weak
  measurement}\cite{AV90} of this variable must yield the same value.
The weak measurement (which is a standard measuring procedure with
weakened coupling) in an appropriate limit does not change the quantum
state of the system. Therefore, a weak spin measurement on one
particle in Hardy's experiment does not affect the state of the
two-particle system and, in particular, the state of the other
particle. Consequently, we can perform weak measurements of spin $x$
components of both particles and the statements $|\Psi\rangle_1 =
|{\downarrow} \rangle_1$ and $|\Psi\rangle_2 = |{\downarrow}
\rangle_2$  remain true. These statements allow us to deduce
the outcomes of these weak measurements, and in  this way our
Lorentz-invariant description tells us the results of actual
experiments.  The limiting operational sense of weak measurements is
due to the fact that usually (and in particular in Hardy's experiment)
an ensemble of pre- and post-selected quantum systems is needed for
obtaining a dispersion-free outcome of a weak
measurement\cite{V-vmer}.
 
It is a pleasure to thank Yakir Aharonov, Lior Goldenberg, and Shmuel
Nussinov  for helpful discussions.  
This research was supported in
part by a grant No. 614/95 from the Israel Science  Foundation.


\end{multicols}


\begin{thebibliography}{99}
\bibitem{foot1}
  Hardy formulates this contradiction in a slightly
  different way: one Lorentz observer concludes that the value of the
  projection operator on the state $|{\downarrow} \rangle_1$ is 1 with
  certainty, ${\bf P}_{|{\downarrow}\rangle_1}=1$, the other concludes
  that ${\bf P}_{|{\downarrow}\rangle_2}=1$, but ${\bf
    P}_{|{\downarrow}\rangle_1}{\bf P}_{|{\downarrow}\rangle_2}=0$.  Note
  that there is no analog of this contradiction in the example of
  three boxes: ${\bf P}_{|B\rangle} +{\bf P}_{|C\rangle} =1$, with
  ${\bf P}_{|C\rangle} =1$ leads to the correct $({\bf P}_{|B\rangle}
  +{\bf P}_{|C\rangle}){\bf P}_{|C\rangle} =1$.
\end{thebibliography}
\end{document}